\documentclass[times,number]{elsarticle}
\usepackage{amsmath}
\usepackage{amssymb}
\usepackage{graphics}
\usepackage{graphicx}
\usepackage{epsfig}
\usepackage{dcolumn}
\usepackage{bm}
\usepackage{lineno}

\begin{document}
\begin{frontmatter}

\title{Plastic scintillator detector array for detection of cosmic ray air shower}

\author[label1]{S.~Roy}
\author[label1]{S.~Chakraborty}
\author[label1]{S.~Chatterjee}
\author[label1]{S.~Biswas\corref{cor}}
\ead{saikat@jcbose.ac.in, saikat.ino@gmail.com, saikat.biswas@cern.ch}
\author[label1]{S.~Das}
\author[label1]{S.~K.~Ghosh}
\author[label1]{A.~Maulik}
\author[label1]{S.~Raha}

\cortext[cor]{Corresponding author}

\address[label1]{Department of Physics and Centre for Astroparticle Physics and Space Science
(CAPSS), Bose Institute, EN-80, Sector V, Kolkata-700091, India}

\begin{abstract}
An air shower array of seven plastic scintillation detectors has been built and commissioned at an altitude of 2200 meter above sea level in the Eastern Himalayas (Darjeeling). Continuous measurement of shower rate using this array is going on since the end of January, 2018. The method of measurement and experimental results are presented in this article.
\end{abstract}
\begin{keyword}
Cosmic ray \sep Shower \sep Scintillator  \sep PMT \sep WLS fibers \sep Astroparticle Physics

\end{keyword}
\end{frontmatter}

\section{Introduction}\label{intro}


Cosmic ray energy spectrum has revealed new fine structures and consensus about the origin and composition of the primary cosmic rays is yet to be reached. Although enormous amount of data has been obtained both through direct measurements by means of satellite or balloon borne experiments as well as indirect such as Extended Air Shower (EAS) experiments \cite{MA90, MA91, KA, WD, SKG9, SKG5}, there exist still a number of open questions on the above mentioned points. One of the reasons for this is the lack of understanding of the hadronic interaction mechanisms during the course of primary cosmic rays through the atmosphere.

At Darjeeling Campus of Bose Institute a hexagonal cosmic ray air shower array has been commissioned in the beginning of 2018. The aim of this program is to study of the properties of cosmic ray at high altitude and also at sea level. In that direction seven plastic scintillator detectors are built and tested successfully. The method of fabrication of detectors are discussed in detail in Ref.~\cite{SB17} and \cite{SR18}. In this article we present the details of the experimental set-up for the shower measurement at high altitude.  


\section{Description of detectors and experimental set-up}\label{setup}

\begin{figure}[htb!]
\begin{center}
\includegraphics[scale=0.5]{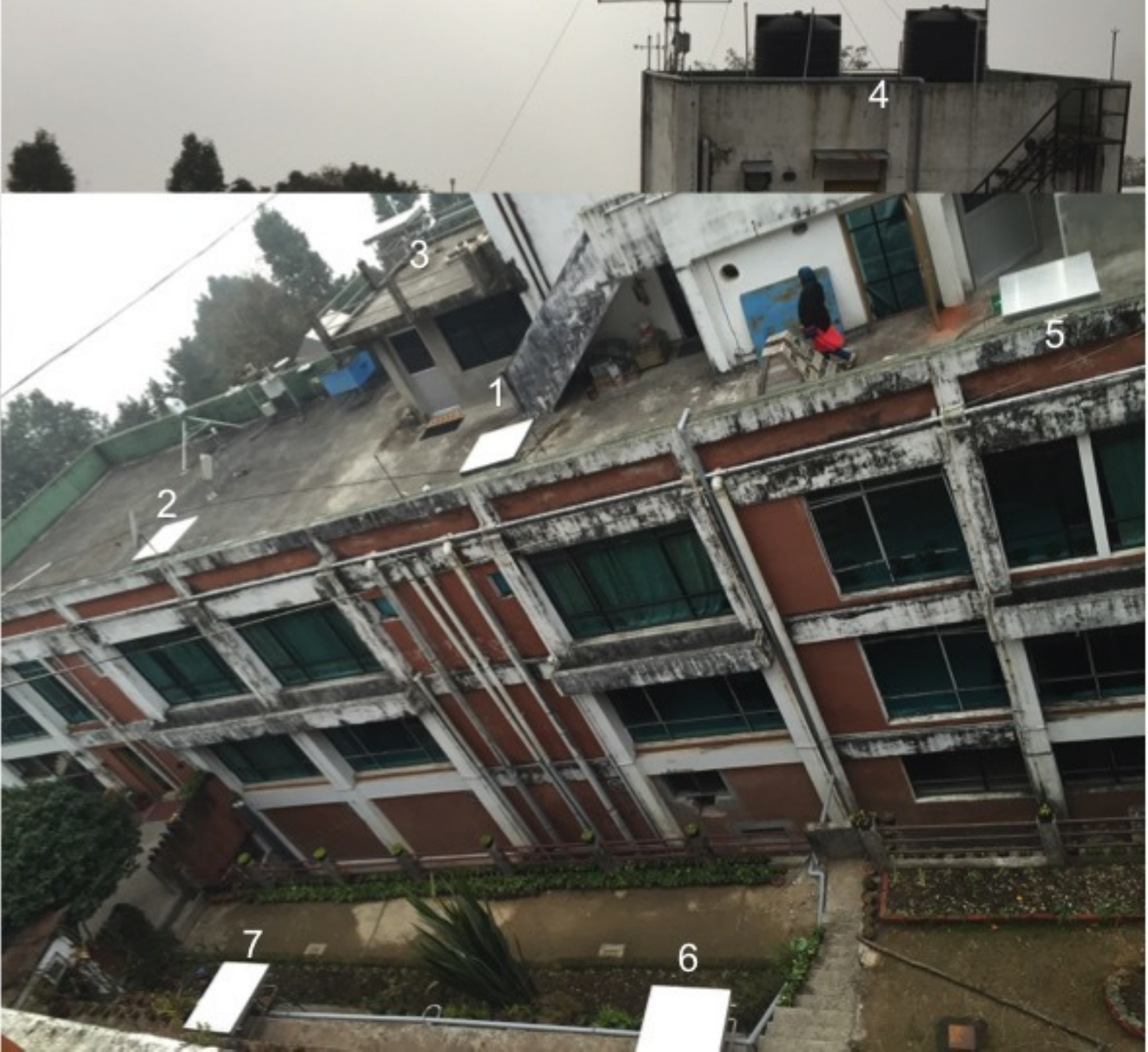}
\caption{Cosmic ray air shower detector array at Darjeeling.}
\label{shower}
 \end{center}
\end{figure}

An array of 7 active detectors have been developed to study cosmic ray extended air showers at an altitude of about 2200 meter above sea level in the Eastern Himalayas (Darjeeling). Six detectors are kept at the vertices of a hexagon and one at the center of it. The distance between any two consecutive detectors is kept to be 8~meter. Each detector element consists of four plastic scintillators of dimension 50~cm~$\times$~50~cm~$\times$~1~cm making the total active area of 1~m~$\times$~1~m. These scintillators have been fabricated indigenously in Cosmic Ray Laboratory (CRL), TIFR, Ooty, India \cite{SKG9, SKG5, PKM9}. All four scintillators of a detector are coupled with a single Photo Multiplier Tube (PMT) using a bundle of 48 wavelength shifting (WLS) fibers spread over the active area of the detector. Each detectors is kept in an Aluminium box and only two connectors, one for the application of high voltage (HV) and one for the signal collection, are fixed to each of them. Each box has been sealed using silicone rubber and kept on a metal stand in open air. The configuration of the cosmic ray air shower detector array at Darjeeling is shown in Figure~\ref{shower}.


From the electronics rack in the counting room, the length of cables for the signal collection are made 50~m for all the elements to eliminate the time delay in signals due to cable length. The length of HV cables are also made 50~m for all detectors. All detectors are biased with -~1740~V from a single HV power supply using an external HV distribution network. The signals from the detectors are fed to a leading edge discriminator (LED) and a common threshold of -~20~mV has been set to all. It has been observed that -~20~mV threshold is enough to eliminate all the noise. A custom-built module with seven inputs has been used to generate multi-fold trigger. Seven individual signals from the discriminator are fed to the trigger module. The shower trigger is generated when the central detector and any two detectors give signal simultaneously. The NIM output from the trigger module has been counted using a scaler module also fabricated at TIFR. The 7-fold coincidence signal from all the scintillator detectors are also counted to measure large shower. The trigger output and the 7-fold signals are counted for 60~minutes to get each data point. 


\section{Experimental results}\label{res}

\begin{figure}[htb!]
\begin{center}
\includegraphics[scale=0.65]{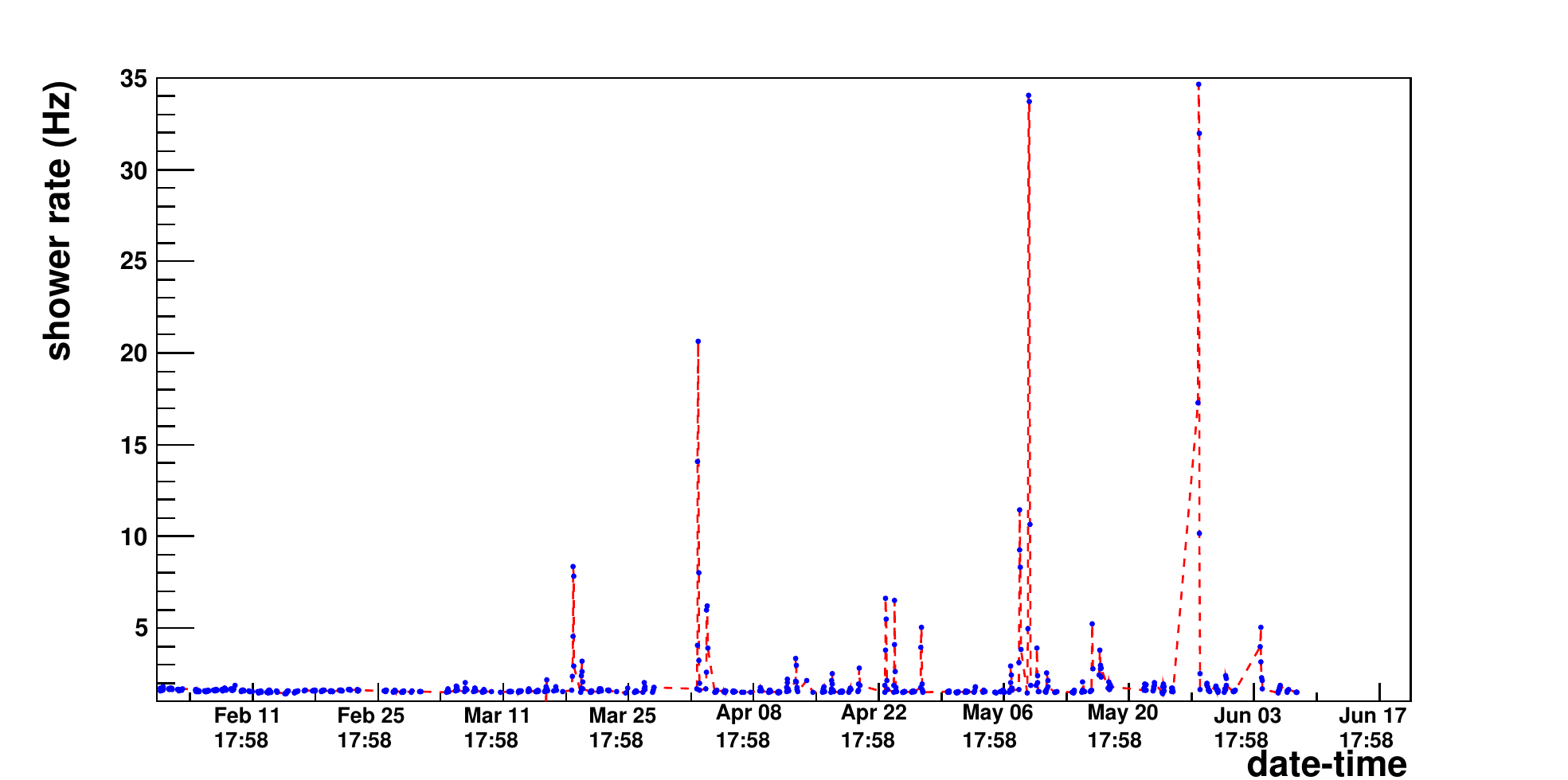}
\caption{Cosmic ray air shower rate as a function of date and time.}
\label{logic}
 \end{center}
\end{figure}
\begin{figure}[htb!]
\begin{center}
\includegraphics[scale=0.65]{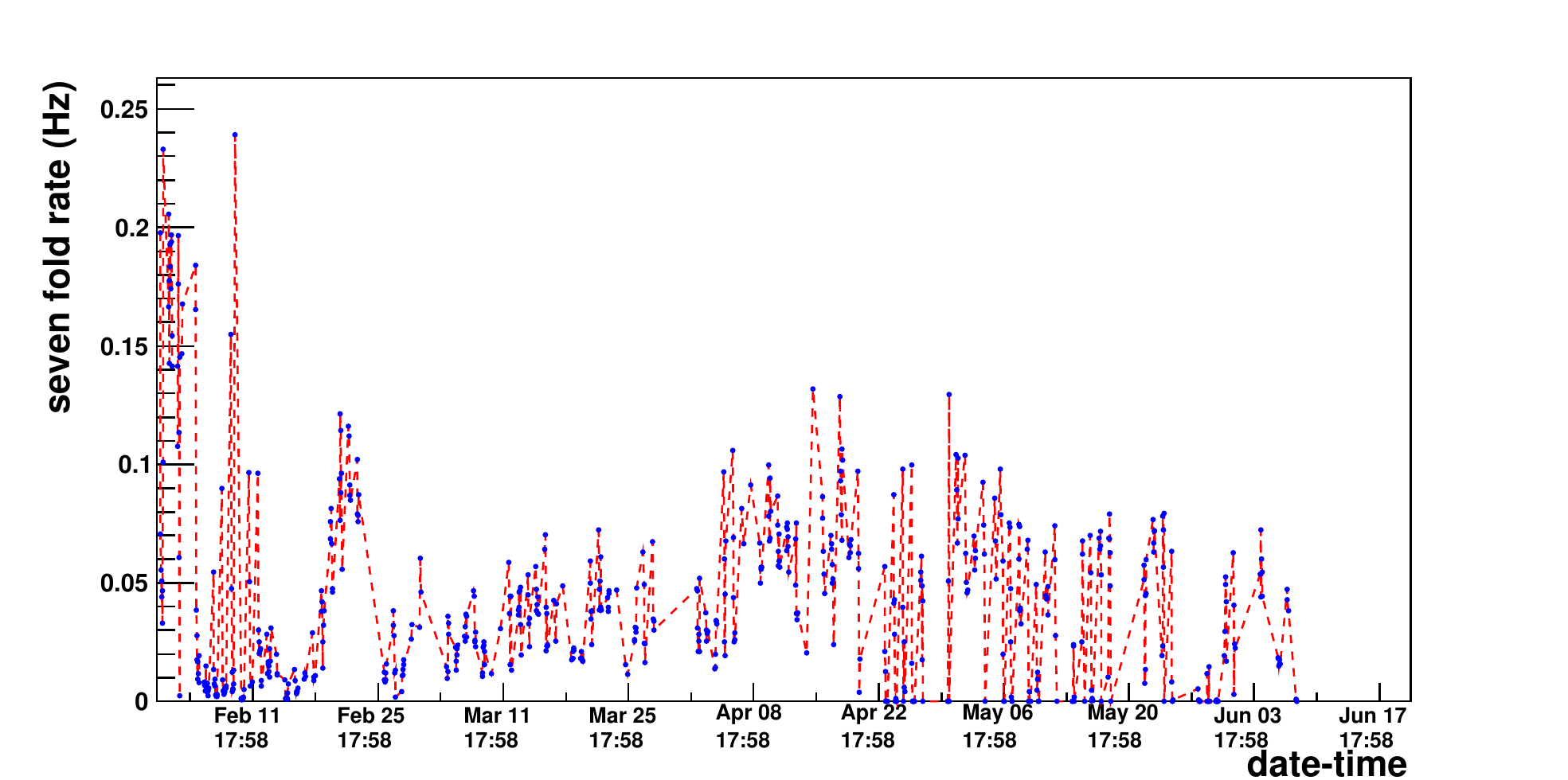}
\caption{7-fold coincidence rate as a function of date and time.}
\label{f7}
 \end{center}
\end{figure}

A continuous measurement of cosmic ray air shower (output of the trigger module) and also the large shower (7-fold coincidence count) are going on since the end of January 2018. The shower rate and the 7-fold coincidence rate as a function of date-time is shown in Figure~\ref{logic} and \ref{f7} respectively. From the measurement (Figure~\ref{logic}) it is observed that the shower rate shows some significantly high values on some days \cite{SF}. It has been noted from other literature that there were some solar flares on some of these days. It is also to be mentioned here that all the spikes obtained are around afternoon (12:00 - 15:00 hours) of those days. However, such behaviour is not present in the measurement of 7-fold coincidence rate (Figure~\ref{f7}).


From a distribution of shower rate it has been found that the average air shower rate, excluding the high values, is $\sim$~1.65~Hz with an RMS of 0.24. During this continuous measurement the 7-fold coincidence rate has been found to be $\sim$~0.04~Hz with an RMS of 0.02. 


\section{Summary}\label{sum}


An array of seven plastic scintillator detectors has been commissioned for detection of cosmic ray air showers at an altitude of about 2200 meter above sea level in the Himalayas at the Centre for Astroparticle Physics \& Space Sciences, Darjeeling campus of Bose Institute. The detector system is continuously measuring the number of cosmic ray showers. From this array it has been found that at an altitude of about 2200 meter the average air shower rate is $\sim$~1.65~Hz with an RMS of 0.24 and the 7-fold coincidence rate has been found to be $\sim$~0.04~Hz with an RMS of 0.02.

So far the detector system has been operated using a NIM electronics system where the data acquisition is done manually. We are now working on integration of the array with a CAMAC based DAQ which will enable recording the data automatically.

After completion of phase I of this project, the objective will be to set up the complete mini-array of 64 such scintillator detectors, at the Darjeeling campus of Bose Institute.


\section*{Acknowledgements}


Authors would like to thank the scientists of Cosmic Ray Laboratory (CRL), Tata Institute of Fundamental Research (TIFR), Ooty, India for fabrication of these scintillators indigenously. We would also like to thank Bose Institute workshop for all the mechanical work. We are thankful to Mrs.~Sumana~Singh for assistance in the assembly of the detector modules. We are also thankful to Mr.~Deb~Kumar~Rai, Mrs.~Yashodhara~Yadav, Mr.~Sabyasachi~Majee, Mr.~Vivek~Gurung for their help in the course of this work. Thanks to Mrs.~Sharmili~Rudra of Seacom Engineering College for some important measurements after building the scintillator detectors at Darjeeling. Finally we acknowledge the IRHPA Phase-II project (IR/S2/PF-01/2011 Dated 26/6/2012) of Department of Science and Technology, Government of India.



\noindent
\begin{thebibliography}{50}

\bibitem{MA90} M. Aglietta et al., Nucl. Phys B \textbf{16} (1990), 493.

\bibitem{MA91} M. Aglietta et al., Europhysics Letters 1 \textbf{15} (1991) 81.

\bibitem{KA} KASCADE Collaboration, Astroparticle Physics \textbf{14} (2001) 245.

\bibitem{WD} W.D. Apel et al., Nucle. Instr. Meth. A \textbf{620} (2010) 202.

\bibitem{SKG9} S.K. Gupta et al., Nuclear Physics B - Proceedings Supplements \textbf{196} (2009) 153M.

\bibitem{SKG5} S.K. Gupta et al., Nuclear Instruments and Methods in Physics Research A \textbf{540} (2005) 311.





\bibitem{SB17} S. Biswas et al., 2017 JINST 12 C06026.  

\bibitem{SR18} S. Roy et al., Proceedings of ADNHEAP 2017, Springer Proceedings in Physics 201, ISBN 978-981-10-7664-0.  



\bibitem{PKM9} P.K. Mohanty et al., Astroparticle Physics \textbf{31} (2009) 24.

\bibitem{SF} https://www.spaceweatherlive.com/en/solar-activity/top-50-solar-flares/year/2018





\end{thebibliography}
\end{document}